\documentclass[a4paper,12pt]{article}
\usepackage{amsmath}
\newcommand{\bm}[1]{\mbox{\boldmath$#1$}}
\title{Is the Thomas precession a source of SR power?}
\author{V.A.~Bordovitsyn\thanks{E-mail: bord@mail.tomsknet.ru.}, 
and A.N.~Myagkii~\thanks{E-mail: myagkii@mail.ru}\\
\it Tomsk State University, Tomsk 634050, Russia}
\date{}

\begin{document}

\begin{titlepage}
\maketitle
\thispagestyle{empty}
\begin{abstract}

The structural composition and the properties of the first quantum
spin-orientation--dependent correction to synchrotron radiation power are
discussed.
On the basis of spin mass renormalization it is shown that,
in the conventional sense, the Thomas precession is not a source of
relativistic radiation. This conclusion is in agreement with well-known
statements on the spin dependence of mass and purely kinematic origin of
Thomas precession.

\end{abstract}



\end{titlepage}

\section{Introduction}

At present, an analysis of the first quantum corrections to the
synchrotron radiation power (SR) is especially topical, because at ultrahigh
energy of electrons in modern accelerators and storage rings, the radiation
effects begin significantly to influence the dynamics and stability of
electron beams.

We consider here polarization and spectral-angular properties of the first
quantum spin-orientation--dependent correction to the synchrotron radiation
power
\begin{equation}
W=W_{SR}(1-\zeta\xi+\ldots),
\label{1}
\end{equation}
where $W_{SR}=\dfrac{2}{3}\dfrac{e_0^2c}{\rho^2}\gamma^4=
\dfrac{2}{3}\dfrac{e_0^2\omega_0^2}{c}\gamma^4$ is the SR power,
$\rho=\dfrac{m_0c^2}{e_0H}\gamma$ is the orbit radius,
$\omega_0=\dfrac{e_0H}{m_0c\gamma}$ is the frequency of electron rotation,
and $\zeta=\pm 1$. the dimensionless parameter $\xi$ can be represented in
different ways:
$$\xi=\frac{3}{2}\frac{\hbar\gamma^2}{m_0c\rho}=
\frac{3}{2}\frac{\hbar\omega_0}{m_0c^2}\gamma^2=
\frac{3}{2}\frac{H}{H^\ast}\gamma=3\frac{\mu_0}{e_0\rho}\gamma^2=
\frac{3}{2}\frac{\hbar}{m_0c^3}\sqrt{w_{\mu}w^{\mu}}=inv,$$
where $H^\ast=\dfrac{m_0^2c^2}{e_0\hbar}$ is Schwinger's magnetic field,
$\mu_0=\dfrac{e_0\hbar}{2m_0c}$ is the Bohr magneton,
$w^{\mu}=\dfrac{d^2r^{\mu}}{d\tau^2}$ is the four-dimensional electron
acceleration, and $e=-e_0>0$ is the electron charge.

The first quantum correction was theoretically calculated by I.M. Ternov,
V.G. Bagrov, and R.A. Rzaev (1964)~\cite{1}. The procedure for experimental
observation of the spin dependence of SR power was proposed by
V.N. Korchuganov, G.N. Kulipanov et al. in 1977, and the experiment itself
was described in detail in~\cite{2}. In 1983, the first quantum
spin-orientation--dependent correction to the SR power was experimentally
detected at the Institute of Nuclear Physics of the Siberian Branch of the
USSR Academy of Sciences (Novosibirsk)~\cite{3}. Later it was found
out~\cite{4} that the correction is not simple in its structural composition.
In the semiclassical theory, it consists of two significantly different
components
$$W_{\rm em}=-\zeta\xi W_{SR}=W_{\rm emL}+W_{\rm emTh},$$
where $W_{\rm emL}$ or $W_{\rm emTh}$ is the spontaneous radiation power
determined by the Larmor or Thomas precession of the electron spin,
respectively. However, the standard classical radiation theory of
relativistic magnetic moment confirms the result only for the Larmor
precession and don not include the contribution of the Thomas precession.
At the same time, all the properties of mixed $\rm emL$-radiation in
classical and semiclassical theory completely coincide~\cite{5}. Here we
try to answer the questions: what is the $\rm emTh$-radiation and why is it
absent in the classical radiation theory of the magnetic moment?

\section{Semiclassical analysis of mixed radiation}

In this section, we use the relativistic semiclassical radiation theory
(Jackson's method~\cite{6}, see also~\cite{7}). In comparison with
the conventional quantum theory of radiation, the method is more simple and
more obvious, but at the same time it reproduces all the results of the
quantum theory.

In the semiclassical theory, the total interaction Hamiltonian has the form
\begin{equation}
\hat{U}^{int}=\hat{U}^{int}_{\rm e}+\hat{U}^{int}_{\rm mL}+
\hat{U}^{int}_{\rm mTh},
\label{2}
\end{equation}
where $\hat{U}^{int}_{\rm e}=e_0(\bm\beta,\tilde{\bm A})$ describes the
interaction of the electron charge with the radiation field via the vector
potential $\tilde{\bm A}$ (ignoring the recoil effects), the other terms
correspond to the interaction of the electron magnetic moment with radiation
fields
\begin{equation}
\hat{U}^{int}_{mL}=\mu\left(\bm\sigma,\left\{\tilde{\bm H}-
[\bm\beta,\tilde{\bm E}]-\frac{\gamma}{\gamma+1}\bm\beta\left(
\bm\beta,\tilde{\bm H}\right)\right\}\right)=
-\frac{\mu}{\gamma}\left\lgroup\bm\sigma,\tilde{\bm H}_0\right\rgroup,
\label{2a}
\end{equation}
\begin{equation}
\hat{U}^{int}_{mTh}=\mu_0\frac{\gamma}{\gamma+1}\left(\bm\sigma,
\left[\bm\beta,\left\{\tilde{\bm E}+
[\bm\beta,\tilde{\bm H}]\right\}\right]\right)=
\frac{\mu_0}{\gamma+1}\left\lgroup[\bm\sigma,\bm\beta],\tilde{\bm E}_0
\right\rgroup.
\label{2b}
\end{equation}

Here $\mu=({g}/{2})\mu_0$ is the total magnetic moment of the electron
including anomalous part, $\bm\sigma$ are the Pauli matrices,
$\bm\beta={\bm u}/{c}$, $\bm u$ is the electron velocity,
$\tilde{\bm E}_0$ and $\tilde{\bm H}_0$ are the radiation fields in
the rest frame of the electron.

It follows from (\ref{2a}) that $\hat{U}^{int}_{\rm mL}$ describes the
interaction of the total magnetic moment with the magnetic field, whereas
$\hat{U}^{int}_{\rm mTh}$ in (\ref{2b}) corresponds to the interaction
induced by the Bohr magneton motion. One can show that in the former case,
the Larmor precession of the spin occurs, whereas in the latter case,
the Thomas precession of the spin takes place. Physically speaking, this
situation is quite obvious, because the Dirac equation involves both
interactions, whereas the anomalous part of the Dirac-Pauli equation involves
only the $\rm mL$-interaction, in other words, the anomalous magnetic moment
undergoes no the Thomas precession.

Calculations of the matrix elements in the semiclassical theory shows that
all the mixed radiation is related to the transitions without a spin flip
(see~\cite{8}). Omitting details of calculations, we write out the spectral
and angular distribution of $\sigma$- and $\pi$- components of mixed
radiation:
$$\frac{dW^{\sigma}_{\rm em}}{dy}=W_{SR}\zeta\xi\cos\nu\frac{9\sqrt{3}}{16\pi}
\left\{\frac{g}{2}\frac{2}{3}y\int^{\infty}_{y}\!\!K_{1/3}(x)dx-2y
\left(\frac{1}{3}\int^{\infty}_{y}\!\!K_{1/3}(x)dx+yK_{1/3}(y)\right)
\right\},$$
\begin{equation}
\frac{dW^{\pi}_{\rm em}}{dy}=W_{SR}\zeta\xi\cos\nu\frac{3\sqrt{3}}{8\pi}
\left(\frac{g}{2}-1\right)y\int^{\infty}_{y}\!\!K_{1/3}(x)dx
\label{3a}
\end{equation}
and
$$\frac{dW^{\sigma}_{\rm em}}{d\chi}=W_{SR}\frac{35}{32}\left\{\frac{g}{2}\chi^2-
(1+\chi^2)\right\}(1+\chi^2)^{-5/2},$$
\begin{equation}
\frac{dW^{\pi}_{\rm em}}{d\chi}=W_{SR}\frac{35}{32}\left(\frac{g}{2}-1\right)
\chi^2(1+\chi^2)^{-9/2}.
\label{3b}
\end{equation}
Here $x=\dfrac{1}{2}(1+\chi^2)^{3/2}y$,
$y=\dfrac{2}{3}\dfrac{\rho\tilde{\omega}}{c\gamma^3}$, $\chi=\gamma\psi$,
$\psi$ is the angle between the direction of radiation and the electron
velocity, $\tilde{\omega}$ is the radiation frequency.

One can obtain the total radiation power by integrating (\ref{3a}) over y
or (\ref{3b}) over $\chi$. Taking into account the main term, which corresponds
to the charge radiation but without the recoil effects, we have
$$W^{\sigma}=W_{SR}\left(\frac{7}{8}+
\zeta\xi\cos\nu\left(\frac{g}{2}-7\right)\frac{1}{6}\right),
\quad W^{\pi}=W_{SR}\left(\frac{1}{8}+
\zeta\xi\cos\nu\left(\frac{g}{2}-1\right)\frac{1}{6}\right),
$$
\begin{equation}
W=W^{\sigma}+W^{\pi}=W_{SR}\left(1+\zeta\xi\cos\nu\left(
\frac{g}{2}-4\right)\frac{1}{3}\right){\left.\right\arrowvert}_{g=2}=W_{SR}\left(1
-\zeta\xi\cos\nu\right).
\label{5}
\end{equation}

In the first quantum correction to the SR power, we specially separate the
terms with the factor ${g}/{2}$ that correspond to the $\rm emL$-radiation
in the correction. The next terms correspond to the $\rm emTh$-radiation.
At $g=2$, we obtain well-known result (\ref{1}) for the Dirac
electron~\cite{1}. It should be noted that at $\nu=\pi/2$ (the spin oriented
in the orbital plane), the mixed radiation is absent.

\section{Classical theory of mixed radiation}

In the classical theory, the mixed radiation is calculated on the basis of the
general radiation theory of the relativistic magnetic moment (see~\cite{8}).
In this section, we use a somewhat different approach.

The energy-momentum density tensor of mixed radiation has the form
$$P^{\mu\rho}_{\rm em}=-\frac{1}{4\pi}\left(\tilde{H}^{\mu\nu}_{\rm e}
\tilde{H}_{\rm m\nu}{}^{\rho}+\tilde{H}^{\mu\nu}_{\rm m}
\tilde{H}_{\rm e\nu}{}^{\rho}+\frac{1}{2}g^{\mu\rho}
\tilde{H}_{\rm e\alpha\beta}\tilde{H}_{\rm m}^{\alpha\beta}\right).$$
Here tensors $\tilde{H}^{\mu\nu}_{\rm e}$ and $\tilde{H}^{\mu\nu}_{\rm m}$
are caused by the radiation field of the charge or the magnetic moment,
respectively:
$$\tilde{H}^{\mu\nu}_{\rm e}=e\left\{-\frac{\tilde{r}^{[\mu}w^{\nu]}}
{{\left(\tilde{r}_{\rho}v^{\rho}\right)}^2}+
\frac{\tilde{r}_{\rho}w^{\rho}\tilde{r}^{[\mu}v^{\nu]}}
{{\left(\tilde{r}_{\rho}v^{\rho}\right)}^3}\right\},$$
$$\tilde{H}^{\mu\nu}_{\rm m}=\mu c\left\{\frac{\stackrel{\circ\circ}
{\Pi}\!{}^{[\mu\lambda}r_{\lambda}\tilde{r}^{\nu]}}{{\left(\tilde{r}_{\rho}
v^{\rho}\right)}^2}-\frac{3\tilde{r}_{\rho}w^{\rho}
\stackrel{\circ}{\Pi}\!{}^{[\mu\lambda}r_{\lambda}\tilde{r}^{\nu]}+
\tilde{r}_{\rho}\stackrel{\circ}{w}\!{}^{\rho}
\Pi^{[\mu\lambda}r_{\lambda}\tilde{r}^{\nu]}}
{{\left(\tilde{r}_{\rho}v^{\rho}\right)}^3}+
3\frac{{\left(\tilde{r}_{\rho}w^{\rho}\right)}^2
\Pi^{[\mu\lambda}r_{\lambda}\tilde{r}^{\nu]}}
{{\left(\tilde{r}_{\rho}v^{\rho}\right)}^4}\right\}.$$
Here $\Pi^{\mu\nu}=\left(\bm\Phi,\bm\Pi\right)$ is the dimensionless
antisymmetric spin tensor which satisfies the condition
$\Pi^{\mu\nu}v_{\nu}=0$ and is related to the Frenkel intrinsic magnetic
moment tensor by the relationship $M^{\mu\nu}=\mu\Pi^{\mu\nu}$,
$\tilde{r}^{\rho}$ is the light-like position four-vector
(charge-observer), $v^{\rho}=dr^{\rho}/d\tau$ is the four-dimensional
velocity, the simbol $\circ$ denotes the proper time derivative.

Substituting these expressions into the four-dimensional momentum of radiation
per unit of proper time
$$\frac{dP^{\mu}_{\rm em}}{d\tau}=\oint P^{\mu\nu}_{\rm em}e_{\nu}d\Omega,$$
where $d\Omega$ is an element of solid angle,
$e^{\nu}=-c\dfrac{\tilde{r}^{\nu}}{\tilde{r}_{\rho}v^{\rho}}-
\dfrac{1}{c}v^{\nu}$ is the unit spacelike four-vector, and integrating over
the angels by means of the well-known method (see~\cite{8}) we obtain
$$\frac{dP^{\mu}_{\rm em}}{d\tau}=\frac{2}{3}\frac{e\mu}{c^4}\left(
\stackrel{\circ\circ}{\Pi}\!{}^{\mu\nu}w_{\nu}-\frac{2}{c^2}v^{\mu}w_{\alpha}
\stackrel{\circ\circ}{\Pi}\!{}^{\alpha\beta}w_{\beta}-\frac{1}{c^2}\Pi^{\mu\nu}
w_{\nu}w_{\rho}w^{\rho}\right).$$
The same result was obtained by another method in~\cite{9} (see
also~\cite{10} and works cited in~{8}). Substituting here the solution
of the spin equation in an uniform magnetic field, we find the mixed
radiation power
($\mu=(g/2)\mu_0$)
$$W_{\rm em}=\frac{c}{\gamma}\frac{dP^0}{d\tau}=-\frac{2}{3}\frac{e\mu}{c^2}
\omega^3\gamma^5\beta^2_{\perp}\Pi_{z}=-\frac{\zeta}{3}W_{SR}\frac{g}{2}.$$

In the case of electron ($e=-e_0$, $\mu=-(g/2)\mu_0$, $\omega=-\omega_0$),
this result with consideration of the main term can be represented in the
form
$$W=W_{SR}\left(1+\frac{1}{3}\zeta\xi\frac{g}{2}\right)=W_{SR}+W_{\rm emL}.$$

We see that the Thomas precession makes no contribution to the total
radiation power. At the same time, all the properties of the
$\rm emL$-radiation in classical and quantum theories completely
coincide~\cite{4,5}.

\section{Physical interpretation of the results obtained}

What is a reason for the discrepancy between the expression of the total
radiation power in classical and quantum (semiclassical) theories?
The situation clears up if we introduce an effective external field
$H^{\mu\nu}_{\rm eff}$. In this case, the equation of spin precession in
the classical theory has especially simple and clear meaning:
\begin{equation}
\frac{d\Pi^{\mu\nu}}{d\tau}=\frac{e}{m_0c}H^{[\mu\rho}_{\rm eff}
\Pi_{\rho}{}^{\nu]},
\label{6}
\end{equation}
\begin{equation}
H^{\mu\rho}_{\rm eff}=H^{\mu\rho}_{\rm L}+H^{\mu\rho}_{\rm Th},\quad
H^{\mu\rho}_{\rm L}=\frac{g}{2}\left(H^{\mu\rho}+\frac{1}{c^2}v^{[\mu}v_{\lambda}
H^{\lambda\rho]}\right)
,\quad
H^{\mu\rho}_{\rm Th}=\frac{m_0}{ec}v^{[\mu}w^{\rho]}.
\label{6a}
\end{equation}

Equation (\ref{6}) may be simplified using the spin vector $\bm\zeta$
specified in the rest frame and related to the components of the tensor
$\Pi^{\mu\nu}$ by means of the Lorentz transformation
$$\Pi^{\mu\nu}=\left\lgroup\gamma[\bm\beta,\bm\zeta],\gamma\bm\zeta-
\frac{\gamma^2}{\gamma+1}\bm\beta\left(\bm\beta,\bm\zeta\right)
\right\rgroup.$$
In this representation, the interpretation of both terms in (\ref{6a})
becomes obvious:
$$\frac{d\bm\zeta}{dt}=[\bm\Omega,\bm\zeta],\quad
\bm\Omega=\bm\Omega_{\rm L}+\bm\Omega_{\rm Th},$$
$$\bm\Omega_{\rm L}=-\frac{eg}{2m_0c}\left(\bm H-[\bm\beta,\bm E]-
\frac{\gamma}{\gamma+1}\bm\beta\left(\bm\beta,\bm H\right)\right)=
-\frac{g}{2}\frac{e}{m_0c\gamma}{\bm H}_0,$$
$$\bm\Omega_{\rm Th}=-\frac{e}{m_0c}\frac{1}{\gamma+1}[\bm\beta,{\bm E}_0]=
-\frac{1}{c}\frac{\gamma^2}{\gamma+1}[\bm\beta,\bm a].$$
Thus, we obtained the well-known expression for the value of the Thomas
precession $\bm\Omega_{\rm Th}$.

It is noteworthy that in the classical theory the interaction of the magnetic
moment with the Thomas field $H^{\mu\nu}_{\rm Th}$ is absent, that is,
$$U^{int}_{\rm mTh}=-\frac{\mu}{2\gamma}H^{\alpha\beta}_{\rm Th}\Pi_{\alpha\beta}=0,$$
whereas in both classical and quantum theories the interaction of the magnetic
moment with the Larmor field assumes absolutely identical forms and hence,
in both theories the interaction has common origin (compare with (2a))
$$U^{int}_{\rm mL}=-\frac{\mu}{2\gamma}H^{\alpha\beta}_{\rm L}
\Pi_{\alpha\beta}=
-\mu\left(\bm\zeta,\left\{\bm H-[\bm\beta,\bm E]-
\frac{\gamma}{\gamma+1}\bm\beta\left(\bm\beta,\bm H\right)\right\}\right)=
-\frac{\mu}{\gamma}\left(\bm\zeta,{\bm H}_0\right).$$

The correspondence principle can be completely understood if we represent
the total radiation power in the semiclassical theory in a somewhat different
manner (compare with formulas (\ref{5}) at $\nu=0$):
$$W^{\sigma}=W_{SR}\left(\frac{7}{8}\left(1-\frac{4}{3}\zeta\xi\right)+
\frac{g}{2}\zeta\xi\frac{1}{6}\right),\quad W^{\pi}=W_{SR}\left(\frac{1}{8}
\left(1-\frac{4}{3}\zeta\xi\right)+\frac{g}{2}\zeta\xi\frac{1}{6}\right),$$
$$W=W_{SR}\left(1-\frac{4}{3}\zeta\xi\right)+W_{\rm emL}.$$
It should be noted that the term $1-\dfrac{4}{3}\zeta\xi$ cannot be associated
with the polarization of radiation. It can be included in $W_{SR}$ at the
expense of spin renormalization of the particle mass (see~\cite{8}
pp.91-93). Indeed, according to the renormalization, the mass of a spin
particle moving in a uniform magnetic field $\bm H=\left(0,0,H\right)$
has the form:
$$M=m_0\left.\left(1-\frac{\mu}{2c}H^{\alpha\beta}\Pi_{\alpha\beta}
\right)\right|_{\mu=-\mu_0}=m_0\left(1+\frac{1}{3}\zeta\xi\right).$$
If we set $E=Mc^2\gamma$, the SR power (\ref{1}) can be represented in the
form
$$W^{\prime}_{SR}=\frac{2}{3}\frac{e^2\omega^2}{c}{\left(\frac{E}{Mc^2}\right)}^4=
W_{SR}\left(1-\frac{4}{3}\zeta\xi\right).$$
Hence we have the relationship
$W=W^{\prime}_{SR}+W_{\rm emL},$
from which it follows that classical and quantum theories of mixed radiation
are in full agreement. Moreover, this means that the
Thomas precession cannot be considered as a source of the SR power.

\end{document}